# Efficient generation of correlated photon pairs in a microstructure fiber

J. Fan[a)] and A. Migdall
*Optical Technology Division,*
*National Institute of Standards and Technology*
*100 Bureau Drive, Mail Stop 8441, Gaithersburg. MD 20899-8441*

L. J. Wang
*Max-Planck Research Group for Optics, Information and Photonics, & Univ. of Erlangen*
*G.-Scharowsky Str. 1, 91058 Erlangen, Germany*

We report efficient generation of correlated photon pairs through degenerate four-wave mixing in microstructure fibers. With 735.7 nm pump pulses producing conjugate signal (688.5 nm) and idler (789.8 nm) photons in a 1.8 m microstructure fiber, we detect photon pairs at a rate of 37.6 kHz with a coincidence/accidental contrast of 10:1 with $\Delta\lambda = 0.7$ nm. This is the highest rate reported to date in a fiber-based photon source. The nonclassicality of this source, as defined by the Zou-Wang-Mandel inequality, is violated by 1100 times the uncertainty.

---

[a)] Electronic address: Jfan@nist.gov

Quantum information science based on the quantum entanglement between multiple parties is fundamentally changing the way we view information and our physical world. Research in these areas has made rapid progress in recent years, although many daunting tasks remain [1-4]. In particular, the practical success of many quantum communication and cryptography applications will require a robust source of correlated photon pairs.

Microstructure fiber (MF), with its central silica core surrounded by patterned air holes, can have very small effective mode diameters (~ 1 to 2 μm) allowing for high field intensities and a wide range of wavelengths (400 nm to 1500 nm) that can propagate as a single spatial mode [5]. These effects greatly increase optical nonlinearities in the fiber, and ease spatial collection. The setup is inherently compatible with telecom operation and free-space operation collection efficiencies can exceed 90%. Four-wave mixing (FWM) in the MF is starting to be considered as an alternative source to provide correlated photons [6-9] beside the well-accepted method of parametric down conversion (PDC) [10-14].

Degenerate FWM and reversed degenerate FWM schemes have been demonstrated in MF to generate correlated photon pairs [6-9]. By injecting a single optical pump wavelength into MF in the negative or zero dispersion spectral region of the fiber, two photons ($\omega_P$) from the pump field are absorbed to create a pair of signal ($\omega_s$) and idler photons ($\omega_i$) under the FWM condition $\omega_s + \omega_i = 2\omega_P$. The contrast between coincidence and accidental coincidence rates (C/A) of 2:1 have been demonstrated [6,7]. Moving the optical pump to the normal dispersion spectral region of the fiber has improved the C/A contrasts to 5:1 [8]. Recently we proposed a conjugate pump scheme that is a reversed degenerate FWM process. A pair of pump beams at conjugate frequencies, with respect to the zero dispersion frequency of the fiber, are the inputs. One photon from each of the two pump beams is absorbed to create a correlated photon pair at the middle frequency under the FWM condition [9]. In that experiment, the highest C/A contrast measured was 8:1.

For high-speed and high-fidelity quantum communication and cryptography applications, two basic conditions are required for the correlated photon source - a high coincidence rate and a high C/A contrast. Here, we report significant advances in each of these parameters using a fiber-based correlated photon source and a degenerate FWM scheme. With signal and idler photons separated in wavelength by 100 nm, we achieve a coincidence rate of 37.6 kHz with a C/A contrast of 10:1 with $\Delta\lambda = 0.7$ nm in free space operation. This is the highest measured rate per bandwidth in a fiber-based photon source to date.

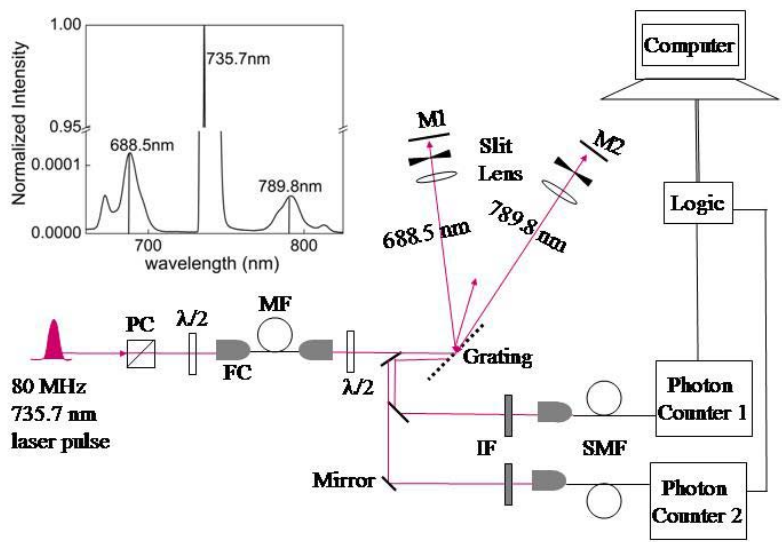

FIG. 1. Schematic experimental setup. PC: polarizer, FC: fiber coupler, BS: beam splitter, SMF: single mode fiber, $\lambda/2$: half wave plate, IF: interference filters. M1 and M2 are mirrors. MF: microstructure fiber. The inset is a normalized spectrum of the output from the MF at an average pump power of 12 mW.

As shown in Fig. 1, we couple linearly polarized laser pulses from a Ti:Sapphire oscillator into a 1.8 m MF (Thorlabs[*]: NL-2.0-735), with polarization along one of the principal axes of the MF.

---

[*] Certain trade names and company products are mentioned in the text or identified in an illustration in order to

The laser wavelength is 735.7 nm (with $\Delta\lambda$ = 0.1 nm), which is the zero dispersion wavelength of the MF. The broadband output from the MF is directed to a high efficiency grating (1,800 lines/mm). A two-pass grating geometry is used to spectrally spread, select, and then re-image the beam to achieve efficient single mode collection. The selected signal and idler wavelengths are $\lambda_s$ = 688.5 nm and $\lambda_i$ = 789.8 nm, each with $\Delta\lambda$ = 0.7 nm set by an adjustable slit. Interference filters ($\Delta\lambda$ = 10 nm) at the collection lenses suppress scattered pump and stray light. The signal and idler photons are coupled into single mode fibers and detected by photon counters and a coincidence circuit, with overall detection efficiencies experimentally determined to be $\eta_s$ = 0.097 for signal photons and $\eta_i$ = 0.076 for idler photons. In the experiment, photon generation rates in the signal and the idler bandwidths are kept below 0.1 photons per pulse.

In MF, the phase matching condition for FWM is $(2k_p - k_s - k_i) - 2\gamma P/(R\tau) = 0$ [15], where $k_p$, $k_s$, and $k_i$ are longitudinal wavevectors for the pump, signal and idler photons in the fiber, $\gamma$ = 110/km/W is the nonlinearity coefficient of the fiber [16], $P$ is the average pump power, $\tau$ = 8 ps is the pump pulse width, and $R$ = 80 MHz is the laser repetition rate. With $P$ = 0.5 mW the detected C/A contrast strongly peaks at $\lambda_s$ = 789.8 nm versus the signal wavelength (Fig. 2(a)), giving $\Delta\lambda$ = 0.9 nm. This correlated phase matching result is the signature of FWM in a MF.

For fixed pump and idler wavelengths, a variation of pump power $\delta P$ = 1 mW produces a wavelength variation of phase-matched signal photons $\delta\lambda_s = \gamma\lambda_s^2 \delta P/(n\pi R\tau) \sim 0.00002$ nm, which is negligible compared to our collection bandwidths. Here $n$ is the refractive index at $\lambda_s$ in the MF. Thus the power-dependent measurement of photon coincidence can be carried out without changing the collection wavelengths.

Fig. 2(b) shows the signal and idler count rates as a function of $P$, both exhibiting strong quadratic dependence with pump power. A maximum coincidence rate of 37.6 kHz with a C/A contrast of 10:1 with $\Delta\lambda = 0.7$ nm at $P = 1$ mW is shown in Fig. 2(c). This exceeds the recent record rate of 6.8 kHz with $\Delta\lambda = 5$ nm to 10 nm at $P = 100$ mW [8]. This is the highest coincidence rate demonstrated to date with a fiber-based correlated photon source.

We attribute this high coincidence and high contrast to several advantages of our scheme. First, it is known that the gains of FWM (and parametric amplification) and spontaneous (and stimulated) Raman scattering are not uniform with wavelength, so it is possible to optimize the relative FWM to Raman signals. In our experiment, this corresponds to placing the signal and idler wavelengths in the high gain spectral region of FWM and parametric amplification shown in the inset of Fig.1, or namely the gain spectral region of the first order Stokes and anti-Stokes. The signal and idler photons are separated by 100 nm in wavelength and are away from the peak-Raman gain wavelength which is 13 THz down-shifted from the pump wavelength in an optical fiber (here it is ~ 760 nm) [15]. This wavelength arrangement can simultaneously maximize the photon pair production rate and the C/A contrast in the MF. A second advantage comes from higher nonlinearity due to both a higher nonlinear coefficient ($\gamma = 110$/km/W) and a higher intensity resulting from the small effective mode diameter (1.2 µm) of the MF used in the experiment. Lastly [16], the two-pass grating arrangement retains the single spatial mode for the signal and idler photons, which improves collection efficiencies. These three advantages together make possible our MF-based photon source's high correlated pair detection rate and high C/A contrast relative to other previous experiments [6-9].

The present result (53.7 kHz/mW/nm) also exceeds coincidence rates by a PDC at similar power levels. Compare this to the 0.2 kHz/mW/nm single–mode PDC coincidence rate reported by Kurtsiefer *et al* [14]. (They reported a 360 kHz rate with $\Delta\lambda$ = 4 nm, but required $P$ = 465 mW.) At a lower pump power of 0.6 mW, the coincidence rate of our fiber-photon source decreases to 10 kHz, but with a higher contrast of 23:1. The highest contrast of 300:1 is measured at 50 μW with a coincidence rate of 45 Hz.

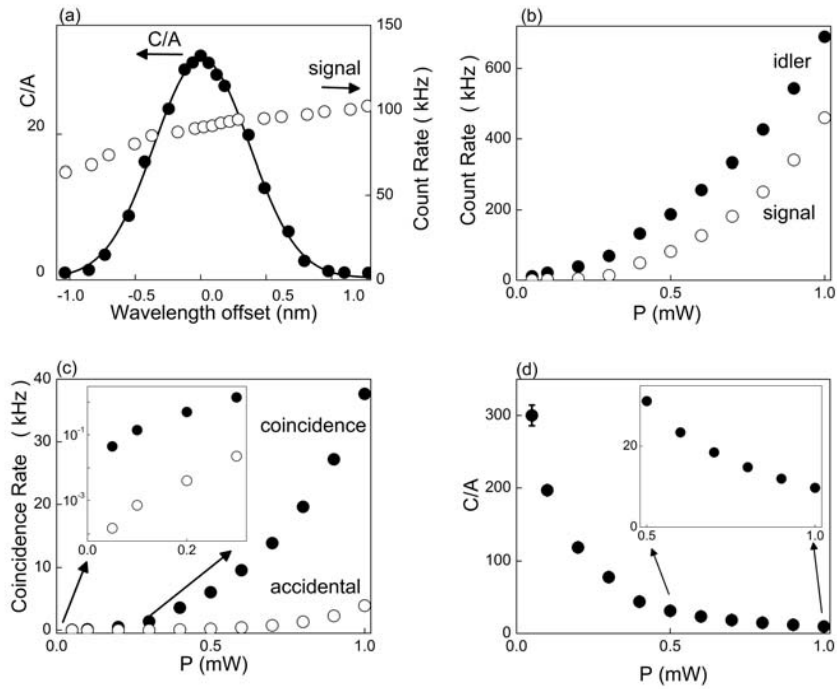

FIG. 2. (a) Spectral scan of C/A (filled dots) and signal photon count rate (open dots) as the signal channel is tuned around the central wavelength of 688.5 nm (corresponding to zero wavelength offset) with pump and idler photons fixed at 735.7 nm and 789.8 nm, respectively, with $P$ = 0.5 mW. C/A is fitted to a Gaussian function (line). (b) Signal (opened dots) and idler photon count rates (filled dots). (c) Detected coincident rate (filled dots) and accidental coincidence rate (open dots). (d) Contrast between the measured coincidence and accidentals. For (a) to (d), each data point is averaged over 30 s for $P \geq 0.4$ mW and 600 s for $P < 0.4$ mW.

The contribution of FWM to photon generation in the MF can be characterized by the ratios of the photon pair generation rate to the signal or idler photon generation rates, defined by $R_s = (D_c - D_a)/(\eta_i D_s)$ and $R_i = (D_c - D_a)/(\eta_s D_i)$, where $D_c$ and $D_a$ are measured coincidence and accidental rates, $D_s$ and $D_i$ are measured signal and idler rates. In our experiment we find $R_s = 96\%$ and $R_i = 50\%$ at $P = 1$ mW and $R_s = 58\%$ and $R_i = 4\%$ at 50 µW, showing that FWM dominates photon generation at $\lambda_s$, while Raman scattering dominates photon generation at $\lambda_i$ and especially at the lower power. The domination of FWM in photon generation at $\lambda_s$ exhibits a high coincidence rate shown in Fig. 2(c) and a high contrast C/A shown in Fig. 2(d).

The observed high coincidence rates cannot be explained with the simple FWM theory developed for a cw optical pump using slowly varying amplitude approximation [17]. The theory predicts a coincidence rate of $\eta |\gamma P z|^2 \Delta\nu \tau R \sim 160$ kHz at $P = 1$ mW which is about 4 times the measured coincidence rate. Here $z$ is the fiber length, $\Delta\nu$ is the detection bandwidth, and $\eta = 0.0074$ is the detection efficiency for a photon pair.

To further examine the photon generation process, we split signal (or idler) photons into two equal beams and examine photon coincidences between them. Self-correlations characterized by the contrast C/A exceed 1 for both signal and idler photons as shown in Fig. 3(a). This shows that a stimulated Raman process and/or a parametric amplification process are involved in photon generation in the MF. These are likely contributors to the disagreement between theory and experiment. The delay induced by the group velocity dispersion between signal and idler photons after propagating 1.8 m in the MF is 2 ps, which is much shorter than the 8 ps width of our pump pulse and can be neglected [16].

If the MF is a nonclassical photon source, the coincidence count rate resulting from the cross-correlation between signal and idler photons should be greater than twice the sum of the self-correlated coincidence rates of signal photons and idler photons [18]. Defining $V = (D_c - D_a) - 2(D_s - D_{s,a} + D_i - D_{i,a})$, where $D_c$ and $D_a$ are the detected cross-correlated coincidence and accidental rates, and $D_s$ ($D_i$) and $D_{s,a}$ ($D_{i,a}$) are the detected self-correlated coincidence and accidentals, for a classical source, we have the Zou-Wang-Mandel inequality $V < 0$. $V$ is measured and plotted in Fig. 3(b). The error bars represent $\sigma$, the combined standard uncertainty. The nonclassicality violation characterized by the ratio of $V/\sigma$ ranges from 360 $\sigma$ to 1100 $\sigma$.

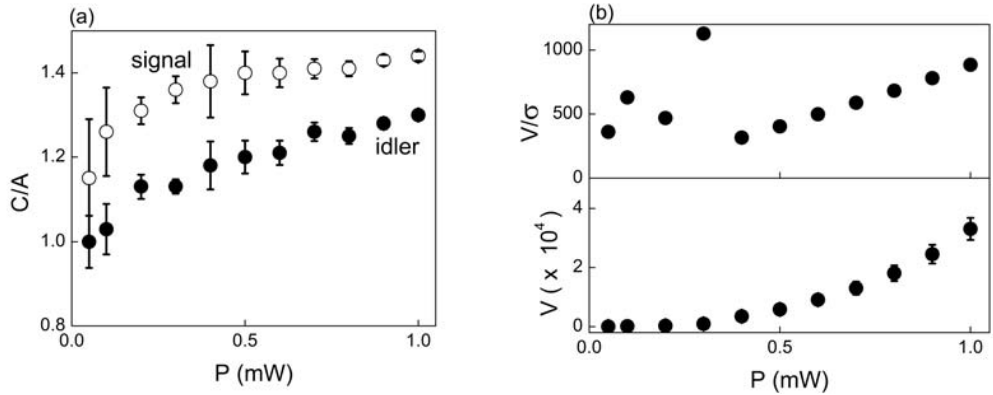

FIG 3. (a) Self-correlation measurements for signal (open dots) and idler photons (filled dots) as a function of the pump power. (b) $V$ (lower) and $V/\sigma$ (upper) vs. $P$, with $V$ error bars expanded by 100x for better view. For (a) and (b), each data point is averaged over 30 s for $P \geq 0.4$ mW and 600 s for $P < 0.4$ mW.

In conclusion, we have experimentally demonstrated the generation of correlated photons in a MF. In a simple system, we have obtained a high twin photon coincidence rate of 53.7 kHz/mW/nm with a contrast of 10:1. This is the highest detection rate of correlated photon pairs in a single mode fiber-based photon source scheme and is ~270 times higher than rates demonstrated using PDC. The classical limit is violated by up to 1100 $\sigma$. We also show that a coincidence/accidental contrast, as high as, 300:1 can be achieved, *albeit* at lower count rates. These high contrasts may be particularly useful in some fundamental tests of quantum mechanics. Our experiment shows that a longer fiber, instead of higher pump power, is the preferred way to improve generation efficiency while suppressing the parametric amplification process. However, group velocity dispersion induced phase mismatch can become a limiting factor in a longer fiber. Our experiment shows that a more complete quantum theory including the contributions from FWM, stimulated Raman scattering, parametric amplification, and group velocity dispersion needs to be developed to model the process of correlated photon generation with a ps pump pulse in MF. Our experiment strongly suggests a practical polarization-entangled correlated photon source can be made with MF and that is being pursued.

Note, since writing this manuscript, it has come to our attention that a similar effort with similar results by Fulconis, et al., has been published [19].

This work has been supported by the MURI Center for Photonic Quantum Information Systems (ARO/ARDA program DAAD19-03-1-0199) and the DARPA/Quist program.